\documentclass[twocolumn,reprint,
 superscriptaddress, amsmath,amssymb, aps, prl]{revtex4-1}
\usepackage{graphicx}
\usepackage{amsmath}
\usepackage{color}
\usepackage{dcolumn}
\usepackage{bm}
\usepackage[
   bookmarks=true,
   colorlinks=true,
   hyperindex=true,
   citecolor=blue,
   linkcolor=blue,
   urlcolor=blue
]{hyperref}

\begin{document}

\title{Balanced Coherence Times of Mixed-Species Atomic Qubits in a Dual $3\times3$ Magic-Intensity Optical Dipole Trap Array}
\author{Ruijun Guo}
\affiliation{State Key Laboratory of Magnetic Resonance and Atomic and Molecular Physics, Wuhan
Institute of Physics and Mathematics, Chinese Academy of Sciences - Wuhan National
Laboratory for Optoelectronics, Wuhan 430071, China}
\affiliation{School of Physics, University of Chinese Academy of Sciences, Beijing 100049, China}
\affiliation{Center for Cold Atom Physics, Chinese Academy of Sciences, Wuhan 430071, China}
\author{Xiaodong He}
\email{hexd@wipm.ac.cn}
\affiliation{State Key Laboratory of Magnetic Resonance and Atomic and Molecular Physics, Wuhan
Institute of Physics and Mathematics, Chinese Academy of Sciences - Wuhan National
Laboratory for Optoelectronics, Wuhan 430071, China}
\affiliation{Center for Cold Atom Physics, Chinese Academy of Sciences, Wuhan 430071, China}
\author{Cheng Sheng}
\affiliation{State Key Laboratory of Magnetic Resonance and Atomic and Molecular Physics, Wuhan
Institute of Physics and Mathematics, Chinese Academy of Sciences - Wuhan National
Laboratory for Optoelectronics, Wuhan 430071, China}
\affiliation{Center for Cold Atom Physics, Chinese Academy of Sciences, Wuhan 430071, China}
\author{Jiaheng Yang}
\affiliation{School of Mathematics and Physics, Qingdao University of Science and Technology, Qingdao 266061, China}
\author{Peng Xu}
\affiliation{State Key Laboratory of Magnetic Resonance and Atomic and Molecular Physics, Wuhan
Institute of Physics and Mathematics, Chinese Academy of Sciences - Wuhan National
Laboratory for Optoelectronics, Wuhan 430071, China}
\affiliation{Center for Cold Atom Physics, Chinese Academy of Sciences, Wuhan 430071, China}

\author{Kunpeng Wang}
\affiliation{State Key Laboratory of Magnetic Resonance and Atomic and Molecular Physics, Wuhan
Institute of Physics and Mathematics, Chinese Academy of Sciences - Wuhan National
Laboratory for Optoelectronics, Wuhan 430071, China}
\affiliation{School of Physics, University of Chinese Academy of Sciences, Beijing 100049, China}
\affiliation{Center for Cold Atom Physics, Chinese Academy of Sciences, Wuhan 430071, China}
\author{Jiaqi Zhong}
\affiliation{State Key Laboratory of Magnetic Resonance and Atomic and Molecular Physics, Wuhan
Institute of Physics and Mathematics, Chinese Academy of Sciences - Wuhan National
Laboratory for Optoelectronics, Wuhan 430071, China}
\affiliation{Center for Cold Atom Physics, Chinese Academy of Sciences, Wuhan 430071, China}
\author{Min Liu}
\affiliation{State Key Laboratory of Magnetic Resonance and Atomic and Molecular Physics, Wuhan
Institute of Physics and Mathematics, Chinese Academy of Sciences - Wuhan National
Laboratory for Optoelectronics, Wuhan 430071, China}
\affiliation{Center for Cold Atom Physics, Chinese Academy of Sciences, Wuhan 430071, China}
\author{Jin Wang}
\affiliation{State Key Laboratory of Magnetic Resonance and Atomic and Molecular Physics, Wuhan
Institute of Physics and Mathematics, Chinese Academy of Sciences - Wuhan National
Laboratory for Optoelectronics, Wuhan 430071, China}
\affiliation{Center for Cold Atom Physics, Chinese Academy of Sciences, Wuhan 430071, China}
\author{Mingsheng Zhan}
\email{mszhan@wipm.ac.cn}
\affiliation{State Key Laboratory of Magnetic Resonance and Atomic and Molecular Physics, Wuhan
Institute of Physics and Mathematics, Chinese Academy of Sciences - Wuhan National
Laboratory for Optoelectronics, Wuhan 430071, China}
\affiliation{Center for Cold Atom Physics, Chinese Academy of Sciences, Wuhan 430071, China}


\begin{abstract}

In this work, we construct a polarization-mediated magic-intensity (MI) optical dipole trap (ODT) array, in which the detrimental effects of light shifts on the mixed-species qubits are efficiently mitigated so that the coherence times of the mixed-species qubits are both substantially enhanced and balanced for the first time. This mixed-species magic trapping technique relies on the tunability of the coefficient of the third-order cross term and ground state hyperpolarizability, which are inherently dependent on the degree of circular polarization of the trap laser. Experimentally, polarization of the ODT array for $^{85}$Rb qubits is finely adjusted to a definite value so that its working magnetic field required for magic trapping amounts to the one required for magically trapping $^{87}$Rb qubits in another ODT array with fully circular polarization. Ultimately, in such a polarization-mediated MI-ODT array, the coherence times of $^{87}$Rb and $^{85}$Rb qubits are respectively enhanced up to 891$\pm$47 ms and 943$\pm$35 ms. Furthermore, a new source of dephasing effect is revealed, which arises from the noise of the elliptic polarization, and the reduction in corresponding dephasing effect on the $^{85}$Rb qubits is attainable by use of shallow magic intensity. It is anticipated that the novel mixed-species MI-ODT array is a versatile platform for building scalable quantum computers with neutral atoms.
\end{abstract}



\maketitle


Neutral atoms confined in an closely spaced array of optical dipole traps (ODTs) manifest outstanding scalability and thus are being intensively developed for quantum simulation and quantum computation~\cite{Saffman2010,Georgescu2014,Saffman2016,Weiss2017}. Recently, atom-by-atom assemblers of defect-free atomic arrays have been demonstrated to deterministically pack 50 qubits into 1D~\cite{Endres2016}, 2D~\cite{Barredo2016} and more qubits into 3D~\cite{Barredo2018,Kumar2018} spaces with relatively compact spacing between qubits. Based on these advances, a 51-atom quantum simulator has been demonstrated ~\cite{Bernien2017}. Besides, the fidelities of universal single-qubit and two-qubit quantum gates have been steadily improved, for the former ones~\cite{Xia2015,Wang2017}, especially, in a novel what is called magic-intensity (MI) ODT array, the performance of global microwave-driven single-qubit Clifford gates have been significantly improved so that errors per gate bellow $10^{-4}$ ~\cite{Sheng2018}; for the latter ones, error rates of 3 percent in entanglement via Rydberg-blockade have been demonstrated~\cite{Levine2018}. The above achievements are important steps along the path of converting the scalability promise of neutral atoms into reality.

When scaling neutral-atom systems to greater number and density of qubits, the problem of crosstalk shows up, which arises from imperfectly isolated logic operation, state readout and initialization of individual qubits. Taking the readout and initialization for example, both involve scattering large numbers of resonant photons, which increases the probability of stray light causing errors on nearby qubits and leads to undesirable recoil heating of the atoms~\cite{Saffman2016}. The low-crosstalk state readout and logic operation are crucially required for implementing quantum error correction~\cite{Devitt2013,Terhal2015}, which is essential for implementing the fault-tolerant quantum computation and allows us to realize the full potential of large-scale quantum information processing devices~\cite{Steane1996,Knill2005}.

To mitigate crosstalk in the multi-qubits quantum processers, it has been being recognized that one of the working approaches is to extend the same-species memory to mixed-species one such that the resonant transition wavelengths differ substantially from each other and allow the spectral isolation and individual addressing of the qubits~\cite{Beterov2015,Saffman2016}. This is analogous to the mixed-species quantum logic spectroscopy previously demonstrated in trapped ions~\cite{Schmidt2005}. Along this line, the mixed-species controlled-NOT quantum gate with a negligible crosstalk have been demonstrated on two single $^{87}$Rb and single $^{85}$Rb atoms in our previous work~\cite{zeng2017}. Prior to neutral atomic qubits, the implementation of entangling quantum gates on mixed trapped-ion qubits have been demonstrated~\cite{Tan2015,Ballance2015}. Furthermore, very recently, the readout of two-qubit stabilizer operators on an ancillary qubit of a different species, which allows readout without crosstalk to the data qubits has been demonstrated in trapped-ion systems~\cite{Negnevitsky2018}.

While the above achievements have efficiently convinced the promising potentials of mixed-species system for quantum error correction, an extra problem arising from employing another species qubits is that the coherence times of the mixed-species are greatly unbalanced. Hitherto, in the mixed-species trapped-ion systems, the coherence times of the ancillary qubits are typically shorter than the data qubits by a factor of larger than  100~\cite{Tan2015,Ballance2015,Negnevitsky2018}. As a result, the strong decoherence of ancillary qubits causes obvious error in measurement and mixed-species logic gates in the error correction~\cite{Negnevitsky2018}. Beyond the quantum computing, particularly in the field of quantum metrology, the long enough coherence times of both mixed-species atomic qubits are crucial to power the quantum-enhanced measurement~\cite{Giovannetti2004,Escher2011,Giovannetti2011}. In order to reliably store coherent superpositions for periods over which quantum error correction can be implemented, it is thus crucial to build a long-lived mixed-species memory. For neutral atoms, one can deploy the magic intensity (MI) technique~\cite{Yang2016} to efficiently suppress the dephasing due to the well-known differential light shift (DLS)~\cite{Kuhr2005,Yu2013}. This MI trapping relies on by applying a bias magnetic B field along a circularly polarized trapping laser field~\cite{Lundblad2010,Derevianko2010} and taking account of a fourth-order hyperpolarizability induced by the trapping fields~\cite{Carr2016}. The open question is whether the magic trapping can be realized in an optically trapped mixed-species neutral-atom memory. This question is explicitly answered in this Letter.


\begin{figure}[htbp]
\centering
\includegraphics[width=8cm]{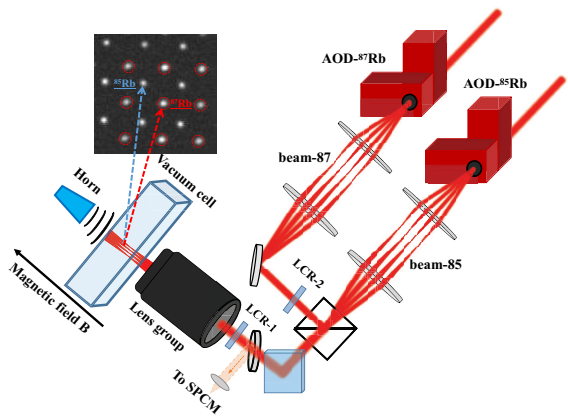}
\caption{(color online). Schematic of the optical layout for building the dual $3\times3$ MI-ODT array. Two 830 nm laser beams from two individual tapered amplifier (one is used for $^{87}$Rb and denoted by beam-87; the other for $^{85}$Rb and denoted by beam-85) are respectively deflected in two orthogonal directions to generate the corresponding 2D beam arrays and then are combined with a non-polarizing 50:50 beamsplitter cube. The dual-axis acousto-optic deflector (AOD) which is driven by a two-channel radio frequency with several tones. The polarizations of beam-87 and beam-85 are engineered via incorporating two liquid crystal retarders (LCRs). An objective lens not only provides a 3D tight confinement but also collects fluorescence from the trapped $^{87}$Rb and $^{85}$Rb atoms. The fluorescence is then counted by a single photon counting module (SPCM). A broadband horn external to the vacuum glass cell is used to deliver the microwave radiation.}
\label{fig:fig1}
\end{figure}

We begin by upgrading the experimental setup for building a 2D ODT array to confine the mixed $^{87}$Rb and $^{85}$Rb qubits, in which the flexible settings of degrees of circular polarization ($\mathcal{A}$) are accessible. Subsequently, we study the behavior of tunability of $B-U_M$ lines for $^{85}$Rb qubits, which are the dependencies of magic intensity $U_M$ on the working magnetic fields B, by tuning the values of $\mathcal{A}$. And we determine the values of $\mathcal{A}$ for magically trapped $^{85}$Rb qubits, in which the required bias B field amounts to the one required for magically trapped $^{87}$Rb qubits. Next we study the coherent characteristic of $^{85}$Rb qubits in such a polarization-mediated MI-ODT array and the accompanying dephasing factors are analysed. Finally, we measure the coherence times of $^{87}$Rb and $^{85}$Rb qubits in a mixed-species polarization-mediated MI-ODT array and find that their coherence times are both enhanced and balanced.

Figure 1 depicts the upgraded experimental setup for producing a 2D mixed-polarization ODT array for building a mixed-species qubits memory. To this end, we use 2 sets of 2-axis acousto-optic deflector (AOD) to deflects two linearly polarized 830 nm laser beams into respective 2D array of beams. Each AOD is driven by 2 channel multi-tone radio-frequency (RF) signals, which is similar to the work of generating a small array of ODT~\cite{Lester2015}. Limited by the availability of 830 nm laser power, we can only make a $3\times3$ ODT array out of each AOD such that trap depth of each site is deep enough to reliably load a single atom from the magneto-optical trap (MOT). The uniformity of the traps are within 12$\%$ after optimization~\cite{Sheng2018}. The 2 sets of output beam arrays are then combined together by a non-polarizing beam splitter. Two pieces of liquid-crystal-retard (LCR) are inserted in such a way that the polarization of the two beams can be individually set to desired polarizations. The resulting beams are then focused to form a 2D mixed-polarization ODT array, in which each spot with a waist of about 1.0 $\mu$m. The sites marked by dashed circles are used to trap $^{87}$Rb atoms, and the unmarked ones are of $^{85}$Rb. The intersite spacings in $^{87}$Rb array and $^{85}$Rb array are both 5.2 $\mu$m.  We note that we use a SPCM instead of electron-multiplied-CCD camera (EMCCD) to detect the fluorescence of individual single atom. Certain single-site detection is realized by scanning the fluorescence image of the array together with spatial filtering techniques.

The mixed-isotope qubits are encoded in microwave clock states of $^{87}$Rb atom as $|0\rangle_{87}\equiv|F=1,m_F=0\rangle$ and $|1\rangle_{87}\equiv|F=2,m_F=0\rangle$ and of $^{85}$Rb atom as $|0\rangle_{85}\equiv|F=2,m_F=0\rangle$ and $|1\rangle_{85}\equiv|F=3,m_F=0\rangle$. The main experimental details on preparation of  $^{87}$Rb and $^{85}$Rb qubits were described in Ref.~\cite{zeng2017}. In brief, we ultimately prepare single $^{87}$Rb and $^{85}$Rb qubits with temperature about 10 $\mu$K and 14 $\mu$K respectively in linearly polarized ODTs with trap depth of 0.4 mK. The coherent properties of the mixed-isotope qubits are studied by the conventional Ramsey interferometry. To do so, the $^{87}$Rb and $^{85}$Rb qubits are respectively rotated by the microwave radiations at frequency about 6.834 GHz and 3.035 GHz. The resonant microwave radiations are delivered via a broadband horn external to the vacuum glass cell.


On the account of degrees of circular polarization $\mathcal{A}$, the DLS of Zeeman-insensitive clock transition seen by the Rb atoms at the external magic field B reads~\cite{Derevianko2010,Carr2016,Yang2016}
\begin{equation}
 \label{eq1}
 \delta\nu(B,U_a,\mathcal{A})=\beta_1 U_a+\mathcal{A}\beta_2 B U_a+\mathcal{A}^2\beta_4 U_a^2
\end{equation}
where $\delta\nu$ is the total DLS seen by the atoms, and $U_a$ (in units
of Hz) is the local light intensity, and $\beta_1$ is the coefficient of third order hyperfine-mediated polarizability, $\beta_2$ is the coefficient of third order cross-term and $\beta_4$ is the coefficient of the groundstate hyperpolarizability. Here, we first determine the corresponding values of the parameters of \{$\beta_1$,$\beta_2$,$\beta_4$\}$_{85}$ for $^{85}$Rb atoms under $\mathcal{A}$=1.00, those are not yet determined. Following the similar approach presented in our previous work~\cite{Yang2016}, under $\mathcal{A}$=1.00, the \{$\beta_1$,$\beta_2$,$\beta_4$\}$_{85}$ are measured to be approximately \{$1.59(4)\times10^{-4},-2.20(8)\times10^{-4}$ G$^{-1}$,$1.09(1)\times10^{-11}$ Hz$^{-1}$\}$_{85}$, those are consistent with the theoretical values \{$1.61(4)\times10^{-4},-2.45(8)\times10^{-4}$ G$^{-1},1.15(1)\times10^{-11}$ Hz$^{-1}$\}$_{85}$, see Ref.~\cite{Derevianko2010,Carr2016}.

\begin{figure}[htbp]
\centering
\includegraphics[width=7cm]{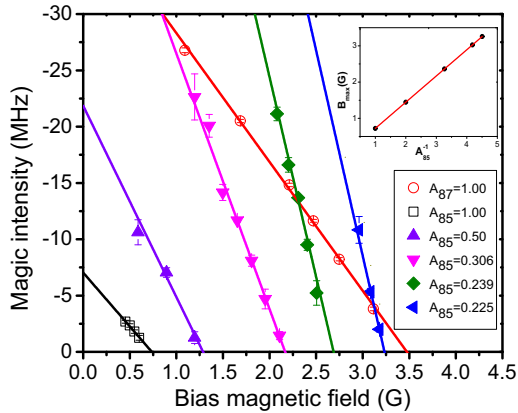}
\caption{(color online). The $\mathcal{A}$-curb on the tunability of $B-U_M$ lines for $^{85}$Rb qubits. Each data point is extracted from the minimum in measured DLS as a function of light intensity (trap depth) for each magnetic B field. The accompanying error bars are fitting errors. For various values of $\mathcal{A}=\{0.500,0.306,0.239,0.225\}$, their corresponding $B-U_M$ lines are recorded, shown respectively as $\{\blacktriangle,\blacktriangledown,\blacklozenge,\blacktriangleleft\}$. The inset shows the measured $B_{max}$ as a function of multiplicative inverse of circular degree, $\mathcal{A}_{85}^{-1}$. }
\label{fig:fig2}
\end{figure}

Obviously, from Eq.(1), the magic trapping happens when $\partial\delta\nu(B,U_a) / \partial U_{a}=0$, yielding the magic intensities $U_M={-(\beta_1+\mathcal{A}\beta_2 B)}{(2\mathcal{A}^2\beta_4)^{-1}}$, those scale linearly with the B fields. This relationship is called $B-U_M$ line. It denotes that the smaller $B$ field sets, the larger intensity requires. But larger intensity means that the atoms scatter more spontaneous Raman photons from the trapping laser, leading to a faster spin relaxation rate. On the other hand, when the working B field reaches a maximum $B_{max} \rightarrow-{\beta_1}/{\mathcal{A}\beta_2}$, which is equal to the $B$-intercept on $B-U_M$ line, $U_M$ approaches 0 and that the trap is too weak to confine atoms. To trade off between the reliable trapping and low spin relaxation rate, the working B field is normally set slightly below the $B_{max}$. Due to about 3.8 GHz gap between hyperfine splitting of $^{87}$Rb and $^{85}$Rb atoms, under the same $\mathcal{A}$, they require respective working $B_{max}$ fields to fulfill the magic operating conditions. For example, when $\mathcal{A}=1.00$, the measured $B_{max}$ for $^{87}$Rb and $^{85}$Rb atoms are respectively 3.50 G and 0.74 G, as shown in Fig.2. The working B field for $^{87}$Rb atoms is typically set to $B_{87}\approx$3.180 G. Obviously, at such magnetic field, no magic depth exists for $^{85}$Rb atoms for the same $\mathcal{A}$.



According to Eq.(1), the $B-U_M$ relationship is tunable as adjusting the $\mathcal{A}$. The $\mathcal{A}$-curb on the local tunability of $B-U_M$ provides a crucially experimental handle on removing the gap of working B filed between mixed-isotope qubits. To demonstrate such tunability, we measure the corresponding $B-U_M$ lines with varied $\mathcal{A}_{85}=\{0.500,0.306,0.239,0.225\}$ for $^{85}$Rb qubits, as shown in Fig.2. The extracted $B_{max}$ is proved to be proportional to the $\mathcal{A}_{85}^{-1}$, as shown in the inset of Fig.2. The clear linear dependence expresses the physics that the coefficient of third-order cross term $\beta_2$ increases linearly with the $\mathcal{A}$ and the insensitivity of the $\beta_1$ to the $\mathcal{A}$. Evidently from the Fig.2, the $B_{max}$ on $B-U_M$ lines of $^{85}$Rb qubits moves towards the one of $^{87}$Rb qubits (open circles) trapping in a MI-ODT with $\mathcal{A}_{87}$ = 1.00 as reducing the $\mathcal{A}_{85}$. 

\begin{figure}[htbp]
\centering
\includegraphics[width=7cm]{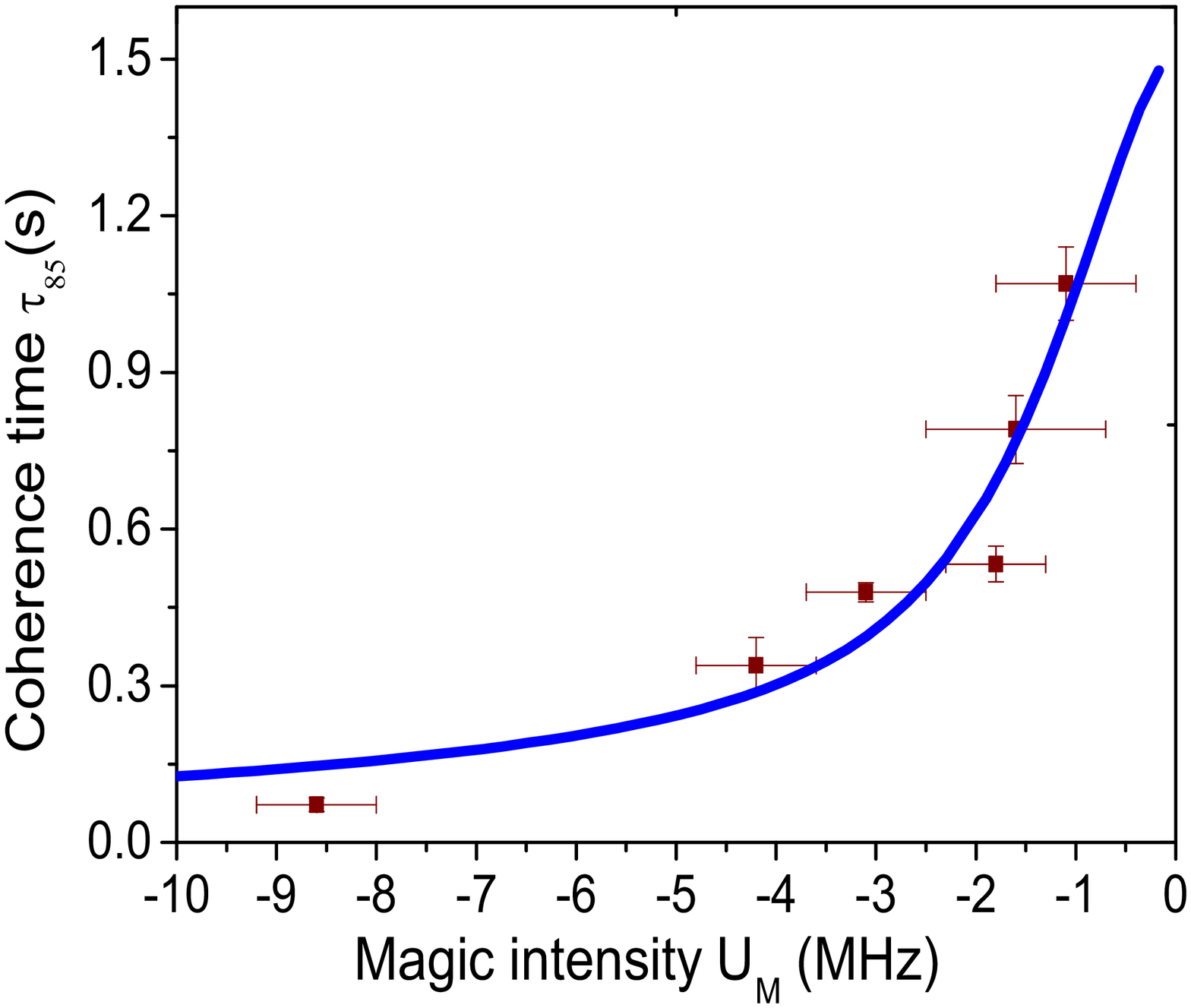}
\caption{(color online). Coherence time of $^{85}$Rb ($\tau_{85}$) qubits and its dependence on the magic intensities ($U_M$) in an elliptically polarized MI-ODT. The solid line is the simulated result via the Monte Calo method on the assumption that the noise of the $\mathcal{A}$ is of Gaussian distribution. All their accompanying error bars of measured $\tau_{85}$ and $U_M$ are of fitting errors.}
\label{fig:fig3}
\end{figure}
Next, we study the coherent property of $^{85}$Rb qubits in an elliptically polarized MI-ODT at the fixed B $\approx$ 3.180 G. To precisely tune the $\mathcal{A}_{85}$ with high resolution, we use the LCR, driven by an AC voltage, to control the polarization. In this work, all the values of $\mathcal{A}_{85}$ are calibrated by fitting the measured intensity-dependent DLS curves to Eq.(1). The coherent property of $^{85}$Rb qubits are measured by the following sequences: First, the polarization of beam-85 is finely tuned so that the resulting $\mathcal{A}_{85}$ is sightly smaller than 0.227; then the corresponding intensity-dependent DLS curve is measured, and the resulting magic intensity is deduced and also the value of $\mathcal{A}_{85}$; eventually, the corresponding coherence time $\tau_{85}$ at the magic intensity is measured via the Ramsey interferometry. The extracted $\tau_{85}$ as a function of magic intensity $U_M$ is plotted in Fig.3. It is evident that the $\tau_{85}$ increases as lowering the magic intensity, and the peak of the coherence time is about 1070 $\pm$70 ms, where the trap depth and the corresponding $\mathcal{A}$ are about -1.0 MHz (48.0 $\mu$K) and 0.225(1) respectively. Those are optimal in practice. Moving to lower trap depths, the MI-ODT is becoming too shallow to reliably confine the $^{85}$Rb atoms since their temperatures are of several $\mu$K.

To analysis the dephasing factors responsible for the measured coherence times, we first employ a fluxgate magnetometer to probe the magnetic noise and determine its contribution to be about $\tau_B\approx$ 1.52 s. It is thus too weak to be responsible for the faster loss of coherence in a MI ODT with a higher trap depth and provides an upper bound for the qubits. On the other hand, both the residual inhomogeneous dephasing arising from the finite temperature of $^{85}$Rb atoms and the longitudinal relaxation time $T_1$ due to spontaneous Raman scattering contribute negligibly to all the involved magic intensities. After excluding the above dephasing factors, the possible one arises out of the noise of $\mathcal{A}$. From Eq.(1), at the magic operating points $U_M$, namely that $\partial{\delta\nu(B,U_a)}/\partial{U_a}\ =\ 0$, the first order of sensitivity of DLS to the noise of $\mathcal{A}_{85}$ is thus $\beta_2 B U_M+2\mathcal{A}_{85}\beta_4 U_M^2$. Obviously sensitivity of the DLS to the noise of $\mathcal{A}$ is quadratically enhanced when the qubits being trapped in a deeper $U_M$. Assuming that the noise of $\mathcal{A}_{85}$ obeys the Gaussian distribution, with mean $\mathcal{A}_{85}$ and the standard derivation $\sigma$. Taking $\sigma$ as a fitting parameter, the dephasing time $\tau_A$ as a function of magic intensity is simulated via the Monte Calo method. By combining $\tau_A$ with $\tau_B$ and $T_1$, the best fit to the data points is as the solid line as plotted in Fig.4. By which, the $\sigma$ is estimated to be 1.0$\times10^{-4}$. It is found out that this noise is mainly from the fluctuating driving voltage of the LCR device. We monitor the change of the driving voltage. The typical peak-to-peak fluctuation is 2$\times10^{-4}$ within 0.7 h.


\begin{figure}[htbp]
\centering
\includegraphics[width=7cm]{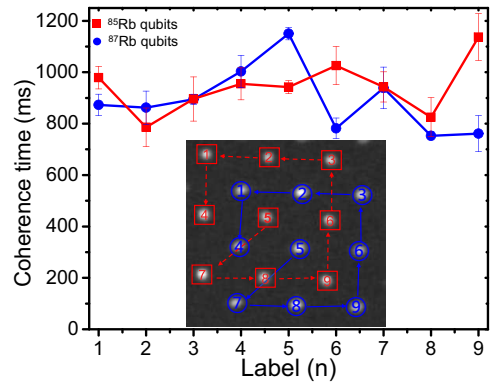}
\caption{(color online). Coherence times of the mixed-isotope qubits. The recorded values of coherence times of $^{87}$Rb (filled circles) and $^{85}$Rb (filled squares) qubits as a function of site number. All the accompanying error bars of coherence times are fitting errors. For the $^{87}$Rb qubits, the lowest and highest coherence times are 753$\pm$16 ms and 1150$\pm$24 ms respectively. While for $^{85}$Rb qubits, the lowest and highest coherence times are 786$\pm$75 ms and 1137$\pm$91 ms respectively. The inset shows the pathes of the scanning sequence of measurement, which are indicated by arrows. Specifically, each site of the array is labeled from 1 to 9, the sites marked with circles and squares are of $^{87}$Rb and $^{85}$Rb respectively, in which the sites marked with 5 are used to calibrated the respective magic operating intensities of arrays.}
\label{fig:fig4}
\end{figure}


At last we measure the coherent properties of $^{87}$Rb and $^{85}$Rb qubits in a 2D polarization-mediated MI-ODT array experimentally. To this end, the $\mathcal{A}_{87}$ of beam-87 remains $1.00$ and the one of beam-85 is finely tuned to be about $\mathcal{A}_{85}\approx$ 0.225. Meanwhile the external magnetic field is set to B $\approx$ 3.180 G. Both sites labeled with number 5 in $^{87}$Rb and $^{85}$Rb arrays, as shown in the inset in Fig.4, are used to calibrated the respective magic operating intensities of arrays. The magic intensity at site 5 for $^{87}$Rb and $^{85}$Rb qubits are respectively set to -2.2 MHz and -1.1 MHz. Since we are using a SPCM for fluorescence detection, we measure the coherence times of mixed-species qubits across the ODT array site-by-site. The sequences of site-by-site scanning are shown in the inset of Fig.4, which are indicated by the arrows. For each site labeled, we carry out Ramsey experiments and obtain its coherence time, as a data point plotted in Fig.4. To ensure the consistency of measuring the coherence times, the laser powers and polarizations of beam-87 and beam-85 unchanged throughout the experiment. After completing experiment, the measured coherence times as a function of site number are plotted in Fig.4. The average coherence time of $^{87}$Rb and $^{85}$Rb qubits are respectively 891$\pm$43 ms and 943$\pm$35 ms. The recorded fluctuations of site-by-site are caused by different factors for different types of qubits. For $^{87}$Rb, it is mainly caused by the inhomogeneity of intensity across the array, since its hyperpolarizability is large so that qubits experience a more homogeneous dephasing effect when the operating intensity is away from the magic operating intensity which is calibrated at site 5, where the longest coherence time appears. For $^{85}$Rb qubits, its sensitivity of coherence time on the inhomogeneity of intensity is abated since the $\mathcal{A}_{85}$ is reduced from unity to 0.225. But the required value of magic intensity is consequentially more sensitive to the drift of $\mathcal{A}_{85}$. Since the temperatures of the LCRs are not stabilized and the measurements take several days long, there are inevitably drifts in response of the LCRs to the light polarizations in these long-term measurements.

In summary, we have devised a polarization-mediated MI-ODT array for mixed-species qubits, in which the detrimental effects of DLS on the mixed-isotope qubits are efficiently mitigated so that the coherence times of the mixed-isotope qubits are substantially enhanced to be close to 1 s and are balanced. Such a long-lived and balanced mixed-isotope memory is a valuable quantum resource for quantum information processing and quantum-enhanced measurements. Our methods demonstrated here can be straightforwardly extended to other optically trapped multi-species neutral atom memories. This work, together with our previous demonstration of entanglement of two individual $^{87}$Rb and $^{85}$Rb atoms via Rydberg blockade~\cite{zeng2017}, represent key steps towards a scalable quantum computer with mixed-species neutral atoms trapped in the polarization-mediated MI-ODT arrays.

\section{acknowledgments}
We acknowledge fruitful discussions with Andrei Derevianko. This work was supported by the National Key Research and Development Program of China under Grant Nos.2017YFA0304501, 2016YFA0302800, and 2016YFA0302002, the National Natural Science Foundation of China under Grant Nos.11774389 and 11704212
, the Strategic Priority Research Program of the Chinese Academy of Sciences under Grant No.XDB21010100 and the Youth Innovation Promotion Association CAS No. 2019325.

\end{document}